\documentclass[pra,12pt,a4paper,amsmath,amssymb,showkeys]{revtex4}

\usepackage{color,epsfig,rotating,graphicx}
\usepackage{amsmath,amssymb,amscd}

\usepackage[latin1]{inputenc}
\usepackage[english]{babel}

\usepackage{dsfont}
\usepackage{textcomp}

\renewcommand{\Re}{{\rm Re}}
\renewcommand{\Im}{{\rm Im}}

\newcommand{\ri}{{\rm i}}

\newcommand{\rd}{{\rm d}}

\newcommand{\kb}{K_{\rm B}}

\begin{document}
\author{S.-A. Biehs}
\email{biehs@theorie.physik.uni-oldenburg.de}
\affiliation{Institut f\"{u}r Physik, Carl von Ossietzky Universit\"{a}t,
D-26111 Oldenburg, Germany.}

\author{G.\ S.\ Agarwal}
\affiliation{Department of Physics, Oklahoma State University, Stillwater, Oklahoma 74078, USA}

\title[\texttt{achemso} FRET]{Large enhancement of F\"{o}rster resonance energy transfer on graphene platforms}


\begin{abstract}
In the view of the applications of F\"{o}rster resonant energy 
transfer (FRET) in biological systems which especially require FRET in 
the inrared region we investigate the great advantage of graphene 
plasmonics in such studies. Focusing on the fundamental aspects of 
FRET between a donor-acceptor pair on a graphene platform showing 
that FRET mediated by the plasmons in graphene is broadband
and enhanced by six orders of magnitude. We briefly discuss
the impact of phonon-polaritonic substrates.
\end{abstract}
\maketitle
\date{\today}
\makeatletter
\begin{center}
\@date
\end{center}
\makeatother

\newpage

%
%

\section{Introduction}

Metal plasmonics has been investigated very intensively in the last decades. It is nowadays well known
that in the visible  metal structures like metal films or nanoparticles, for instance, are the plasmonic 
structures of choice exhibiting strong field confinement and large field enhancement close to the metallic 
structure. In contrast, research on graphene plasmonics which is in some sense complementary to  metal 
plasmonics~\cite{Interview} has just begun. The plasmons in graphene show strong field confinement in the 
infrared, whereas in the visible graphene plasmonics is challenging~\cite{JablanEtAl2009,KoppensEtAl2011,BrarEtAl2013}. 
One of the main advantages of using graphene in plasmonics or graphene-based hybrid-plasmonic devices is its 
tunability by doping or gating~\cite{GrigorenkoEtAl2012}. Especially for applications of the F\"{o}rster resonance energy transfer (FRET) such as {\itshape in vivo} infrared flourescence imaging applications as for example 
lymph-node mapping and cancer imaging~~\cite{HeEtAl2012,XiongEtAl2012} graphene plasmonics is highly suitable and 
can unfold its full strength.

The F\"{o}rster resonant-energy transfer (FRET) itself is widely used in biochemistry to study 
protein and RNA folding~\cite{Weiss2000}, DNA nanomechanical devices~\cite{MuellerEtAl2006} and 
even to transport energy along a DNA backbone~\cite{VyawahareEtAl2004} to mention a few applications. 
Theoretically it was already shown in the 80's that in the vicinity of nanoparticles FRET between a 
donor and an acceptor molecule can be enhanced by two to five orders of magnitude in the idealized case
when both the donor and acceptor are in resonance with the localized plasmon resonance~\cite{GerstenNitzan1984,HuaEtAl1985}. 
Recent theoretical works have also studied the impact of nonlocal effects and the possibility of 
magneto-optical control of FRET close to metallic nanoparticles~\cite{XieEtAl2009,VincentCarminati2011}.
Although the plasmonic enhancement (PE) effect using nanoparticles could be measured lately~\cite{MalickaEtAl2003,ReilEtAl2008}, only 
moderate enhancement factors of 3.5 and 8.6 were found~\cite{ReilEtAl2008}. More recent experiments
report FRET enhancements by a factor of 173~\cite{VigerEtAl2011}. The PE could also be demonstrated experimentally for 
FRET between a donor and acceptor separated by a metal film~\cite{AndrewBarnes2004}.   

\begin{figure}
 \epsfig{file = 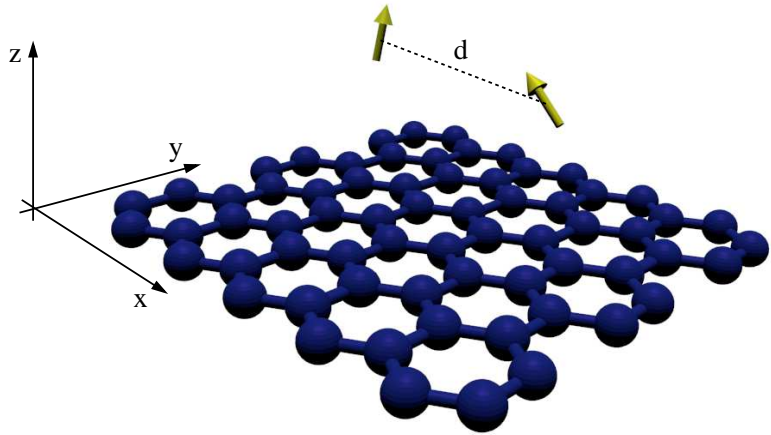, width = 0.7\textwidth}
  \caption{Sketch of the donor and acceptor above a sheet of graphene. \label{Fig:Sketch}} 
\end{figure}

In this Letter, we demonstrate how graphene plasmonics can enhance FRET in the infrared. 
We want to emphasize that most recent studies are considering spontaneous emission close to 
graphene~\cite{KoppensEtAl2011,VelizhaninEfimov2011,TislerEtAl2013,MessinaEtAl2013} which 
is a {\itshape first-order} process, whereas FRET is a {\itshape second-order} process~\cite{AgarwalBook}. 
We study in detail the FRET between a donor-acceptor pair in close vicinity to a graphene sheet as depicted 
in Fig.~\ref{Fig:Sketch}. We show that FRET mediated by the plasmons in the graphene sheet is broadband and can be 
enhanced by {\itshape six} orders of magnitude. The broadband property of the PE is very advantageous since one can expect large
PE also in cases where the emission and absorption spectra of the donor and acceptor do not fully overlap.
Further we show that the presence of graphene allows for energy transfer between perpendicularly
oriented dipole moments which do not couple in free space. Finally, we discuss how the interaction
of the graphene plasmons with the surface phonon polaritons of a SiC substrate affects the energy transfer.
Since graphene is highly tunable this PE effect can be easily controlled even dynamically. 

%
%

\section{Resonant dipole-dipole interaction}

Using second-order perturbation theory the energy transfer rate for dipole-dipole interaction between a 
donor and an acceptor in the presence of arbitrarily shaped, dispersing and absorbing bodies can be 
written as~\cite{DungEtAl2002}
\begin{equation}
  \Gamma = \int \!\!\! \rd \omega\, \sigma^{\rm abs}(\omega) T(\omega) \sigma^{\rm em}(\omega)
\end{equation}
where $\sigma^{\rm abs}$ and $\sigma^{\rm em}$ are the {\itshape free space} absorption and emission 
spectra of the acceptor and the donor. The energy transfer function $T(\omega)$ between the donor and 
acceptor is~\cite{DungEtAl2002} 
\begin{equation}
  T(\omega) = \frac{2 \pi}{\hbar^2} \biggl( \frac{\omega^2}{\epsilon_0 c^2} \biggr)^2 |\mathbf{d}_{\rm D}|^2 |\mathbf{d}_{\rm A}|^2 |\mathbf{e}_{\rm A} \cdot \mathds{G}(\mathbf{r}_A,\mathbf{r}_D) \cdot \mathbf{e}_{\rm D}|^2
\end{equation}
introducing the dipole-transition matrix elements $\mathbf{d}_{\rm D} = |\mathbf{d}_{\rm D}| \mathbf{e}_{\rm D}$ and 
$\mathbf{d}_{\rm A} = |\mathbf{d}_{\rm A}| \mathbf{e}_{\rm A}$ of the donor and acceptor. The unit vectors
$ \mathbf{e}_{\rm D}$ and  $\mathbf{e}_{\rm A}$ determine the orientation of the dipole moments. $\epsilon_0$ is the permittivity 
of vacuum. Through the Green's function  $\mathds{G}(\mathbf{r}_A,\mathbf{r}_D)$ the environment of the donor 
at $\mathbf{r}_D$ and the acceptor at $\mathbf{r}_A$
is fully taken into account. The energy transfer function $T(\omega)$ can also be expressed in terms of
the free space spontaneaous emission rates~\cite{NovotnyHecht}
\begin{equation}
  \gamma_{D/A} = \frac{\omega^3 |\mathbf{d}_{D/A}|}{3 \pi \epsilon_0 \hbar c^3}.
\end{equation}
We obtain
\begin{equation}
  T(\omega) = 18 \pi^3 \gamma_A \gamma_D \frac{ |\mathbf{e}_{\rm A} \cdot \mathds{G}(\mathbf{r}_A,\mathbf{r}_D) \cdot \mathbf{e}_{\rm D}|^2}{(2 \pi / \lambda)^2}.
\end{equation}
By normalizing the emission and absorption spectra by the free space emission rates $\tilde{\sigma}^{\rm em} \equiv \sigma^{\rm em} \gamma_D$ and $\tilde{\sigma}^{\rm abs} \equiv \sigma^{\rm abs} \gamma_A$ we can rewrite the expression for the energy transfer as
\begin{equation}
  \Gamma = \int \!\!\! \rd \omega\, \tilde{\sigma}^{\rm abs}(\omega)\tilde{T}(\omega) \tilde{\sigma}^{\rm em}(\omega)
\end{equation}
introducing the dimensionless energy transfer function
\begin{equation}
  \Tilde{T} (\omega) \equiv 18 \pi^3 \frac{ |\mathbf{e}_{\rm A} \cdot \mathds{G}(\mathbf{r}_A,\mathbf{r}_D) \cdot \mathbf{e}_{\rm D}|^2}{(2 \pi / \lambda)^2}.
\end{equation}

The enhancement factor for the energy transfer close to a plasmonic structure like graphene in comparison to vacuum is defined as
\begin{equation}
  E(\omega) \equiv \frac{\tilde{T}(\omega)}{\tilde{T}^{(\rm vac)} (\omega)}  =  \frac{ |\mathbf{e}_{\rm A} \cdot \mathds{G} \cdot \mathbf{e}_{\rm D}|^2}{ |\mathbf{e}_{\rm A} \cdot \mathds{G}^{(\rm vac)} \cdot \mathbf{e}_{\rm D}|^2}
\label{Eq:enhancementfactor}
\end{equation}
where the Green's function $\mathds{G} = \mathds{G}^{(\rm vac)} + \mathds{G}^{(\rm sc)}$ is a sum of the vacuum 
Green's function $\mathds{G}^{(\rm vac)}$ and the scattered Green's function $\mathds{G}^{(\rm sc)}$. The explicit 
expressions of the Green's functions for free space and planar structures can be found in Refs.~\cite{NovotnyHecht,Sipe} (see also Supporting Information).

%
%

\section{Optical properties of graphene}

Before we can determine the energy transfer rate between the donor and acceptor close to a sheet of graphene it is necessary to
determine its optical properties which enter in the scattered Green's function. The optical properties of graphene are fully 
described by its in-plane conductivity 
\begin{equation}
  \sigma = \sigma_D + \sigma_I
\end{equation}
consisting of the Drude-like intraband contribution $\sigma_D$ and the interband contribution $\sigma_I$ which are 
given by~\cite{FalkovskyVarlamov2007}
\begin{align}
  \sigma_D &= \frac{\ri}{\omega + \ri/\tau} \frac{2 e^2 \kb T}{\pi \hbar^2} \log\biggl[ 2 \cosh\biggl( \frac{E_F}{2 \kb T} \biggr) \biggr], \\
  \sigma_I &= \frac{e^2}{4 \hbar} \biggl[ G\biggl( \frac{\hbar \omega}{2} + \ri \frac{4 \hbar \omega}{\pi} \int_0^\infty \rd \xi \, \frac{G(\xi) - G(\hbar\omega/2)}{(\hbar \omega)^2 - 4 \xi^2} \biggr) \biggr]
\end{align}
where
\begin{equation}
  G(x) = \frac{\sinh(x/\kb T)}{\cosh(E_F/\kb T) + \cosh(x/\kb T)}.
\end{equation} 
Here $T$ is the temperature of the graphene sheet, $E_F$ is its Fermi-level which can be tuned by gating and doping for instance. 
$\tau$ is a phenomenological damping constant, $\kb$ is Boltzmann's constant, $2 \pi \hbar$ is Planck's constant and $e$ is the electron charge. 
The in-plane conductivity enters in the Fresnel reflection coefficients. For a graphene sheet on a dielectric substrate with permittivity 
$\epsilon$ the reflection coefficients for s- and p-polarized light are given by~\cite{StauberEtAl2008,KoppensEtAl2011} (see also Supporting Information)
\begin{align}
  r_{\rm s} &= \frac{\gamma_0 - \gamma - \sigma \omega}{\gamma_0 + \gamma + \sigma \omega}, \\
  r_{\rm p} &= \frac{\epsilon \gamma_0 - \gamma + \frac{\sigma \gamma \gamma_0}{\omega \epsilon_0}}{\epsilon \gamma_0 + \gamma + \frac{\sigma \gamma \gamma_0}{\omega \epsilon_0}},
\end{align}
where $\kappa = (k_x,k_y)^t$ is the in-plane wave vector and $\gamma = \sqrt{k_0^2 \epsilon - \kappa^2}$ and 
$\gamma_0 = \sqrt{k_0^2 - \kappa^2}$ are the out-of-plane wave vectors inside the substrate and in vacuum. 
The substrate can have quite a large impact on the properties
of the plasmons in graphene, in particular, when it supports surface phonon polaritons like SiO$_2$ and SiC~\cite{KochEtAl2012,YanEtAl2013,MessinaEtAl2013}. For convenience we will neglect the influence of the substrate in the following and focus 
mainly on suspended graphene.

%
%

\section{Plasmons in graphene}

The plasmons in graphene propagating along the sheet are determined by the poles of the reflection
coefficient for p polarization. When assuming that we have suspended graphene ($\epsilon = 1$) then
the dispersion relation of the plasmon simply reads
\begin{equation}
  \kappa_{\rm P} = \sqrt{k_0^2 - \biggl(\frac{2 \omega \epsilon_0}{\sigma}\biggr)^2}.
\label{Eq:KappaP}
\end{equation}
From this dispersion relation one can easily determine the wavelength $\lambda_{\rm P} = 2 \pi/\Re(\kappa)$
and the propagation length $l_{\rm P} = 1/\Im(\kappa)$ of the plasmon. For a relatively high Fermi level
$E_{\rm F} = 1\,{\rm eV}$ which can be realized nowadays~\cite{GrigorenkoEtAl2012} Table~\ref{Table1} summarizes
some of the plasmon properties as the in-plane confinement and the propagation length for frequencies ranging
from $0.2-0.9\,{\rm eV}$.
The in-plane confinement $\lambda_{\rm P} / \lambda$ is in this case strong and can attain values of $0.013$ to
$0.072$. The out-of-plane confinement is even stronger: since the out-of-plane wave vector is approximately 
$\ri \kappa_{\rm P}$ the out-of-plane confinement is given by $1/\Re(\kappa_{\rm P}) = \lambda_{\rm P}/(2 \pi)$.
This means that the out-of-plane confinement is between $0.002 \lambda$ and $0.011 \lambda$. The propagation length 
$l_{\rm P}$ of the plasmons in graphene is found to be between $300\,{\rm nm}$
and $2\,\mu{\rm m}$ depending on the frequency so that $l_{\rm P} / \lambda_{\rm P}$ is between $17.4$ and $4.8$.  
Note that the propagation length approximately scales linearly with the relaxation time $\tau$ so 
that $l_{\rm P} / \lambda_{\rm P}$ can even reach values on the order of 100 for $\tau = 10^{-12}\, {\rm s}^{-1}$
as shown in Ref.~\cite{KoppensEtAl2011}. Further note that the plasmon properties can be easily tuned by changing
the Fermi level by gating or tuning.

\begin{table}
  \begin{tabular}{ | c | c | c | c | c |} \hline
    frequency (eV) & $\lambda$ ($\mu m$)&$\lambda_{\rm P} / \lambda$ & $l_{\rm P}$ (nm) & $l_{\rm P} / \lambda_{\rm P} $ \\ \hline
    0.2 & 6.23 & 0.072 & 2142 & 4.8  \\ \hline
    0.3 & 4.15 & 0.047 & 1397 & 7.08 \\ \hline
    0.5 & 2.49 & 0.027 & 773  & 11.3 \\ \hline
    0.7 & 1.78 & 0.018 & 486  & 14.9 \\ \hline
    0.9 & 1.38 & 0.013 & 311  & 17.4 \\ \hline
  \end{tabular}
  \caption{The in-plane confinement $\lambda_{\rm P} / \lambda$, the propagation length $l_{\rm P}$, and $l_{\rm P}/\lambda_{\rm P}$ of the surface plasmons in graphene using $T = 300\,{\rm K}$, $E_F = 1\,{\rm eV}$ and $\tau = 10^{-13} \, {\rm s}^{-1}$.\label{Table1}}
\end{table}

%
%

\section{Large FRET mediated by graphene plasmons}

\begin{figure}
  \epsfig{file = 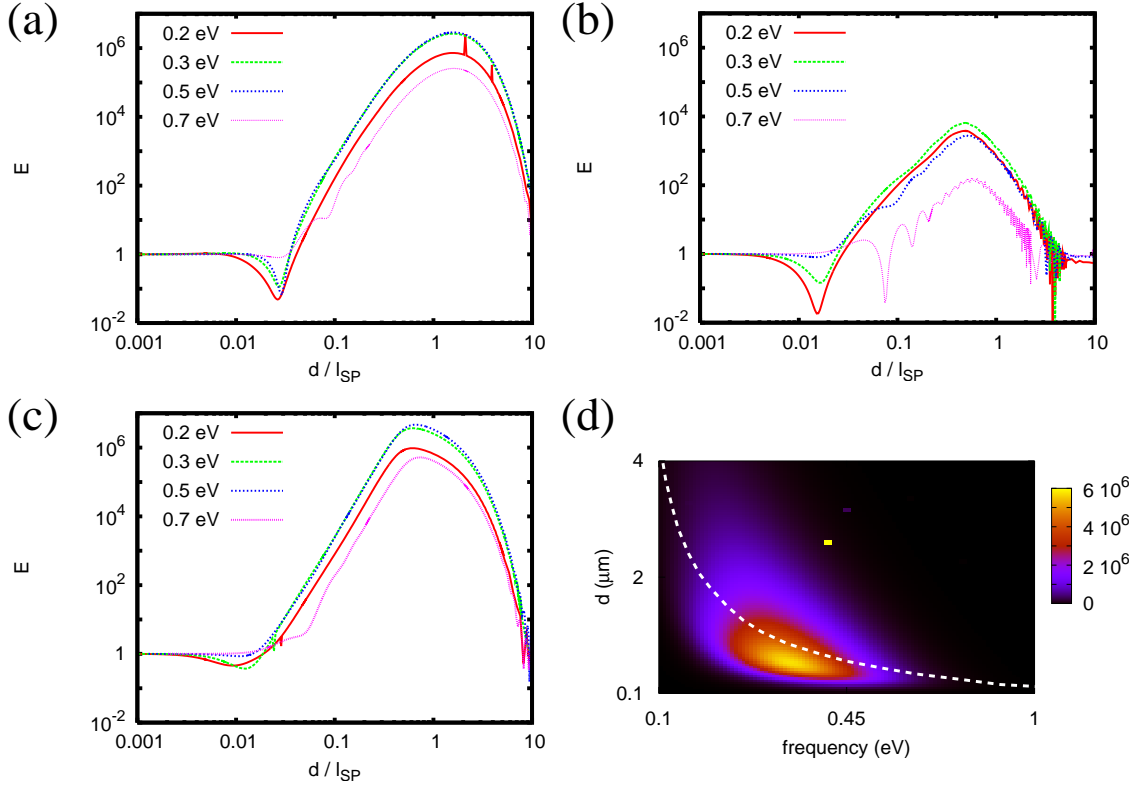, width = 0.9\textwidth}
  \caption{(a)-(c) Enhancement factor $E$ for donor and acceptor placed along the x-axis as a function of distance $d$ for different frequencies ranging from $0.2$ to $0.7\,{\rm eV}$ ($z = 10\,{\rm nm}$ and $E_F = 1\,{\rm eV}$). The distance is for each frequency normalized to the corresponding propagation length $l_{\rm P}$ of the graphene plasmon. The orientations of the dipole moment are (a) $\mathbf{e}_{\rm A} = \mathbf{e}_{\rm D} = \mathbf{e}_x$, (b) $\mathbf{e}_{\rm A} = \mathbf{e}_{\rm D} = \mathbf{e}_y$, and (c) $\mathbf{e}_{\rm A} = \mathbf{e}_{\rm D} = \mathbf{e}_z$. (d) Enhancement factor $E$  for $\mathbf{e}_{\rm A} = \mathbf{e}_{\rm D} = \mathbf{e}_z$ as a function of frequency and distance. The white dashed line is a plot of $1/\Im(\kappa_{\rm S})$ from Eq.~(\ref{Eq:KappaP}). \label{Fig:distancedependence}}
\end{figure}

In order to demonstrate the PE we consider a donor and an acceptor on a plane parallel to the
graphene sheet. The distance of this plane to the sheet of graphene which is assumed
to be in the x-y plane is choosen to be $z = 10\,{\rm nm}$. In Fig.~\ref{Fig:distancedependence}(a)-(c) 
the enhancement factor $E$ defined in Eq.~\ref{Eq:enhancementfactor} is plotted as a function of distance $d$
between the donor and the acceptor for parallely aligned dipole moments along the $x$, $y$ and
$z$ axis. It can be seen that FRET mediated by the plasmons in graphene can be {\itshape six} orders of magnitude
larger than in free space. That means that also the recently studied nonparaxial spin-Hall effect of light~\cite{AgarwalBiehs2013} 
will be extraordinary large in the vicinity of graphene. Further the maximum enhancement is obtained for distances on the order
of the propagation length of the graphene plasmons. In Fig.~\ref{Fig:distancedependence}(d) we show
a plot of the enhancement factor $E$ as a function of distance $d$ and frequency $\omega$ for
the case where the dipole moments of the donor and acceptor are oriented along the z axis. It becomes
apparent that the PE is broadband in frequency for distances $d$ between $100\,{\rm nm}$ and $1\,\mu{ \rm m}$. Hence,
one can expect that the predicted large enhancement of FRET can be achieved in the realistic case where the emission 
and absorption spectra of the donor and acceptor are nonresonant in contrast to the case of metallic nanoparticles where
the resonance of the localized plasmon is narrow band. 

\begin{figure}
  \epsfig{file = 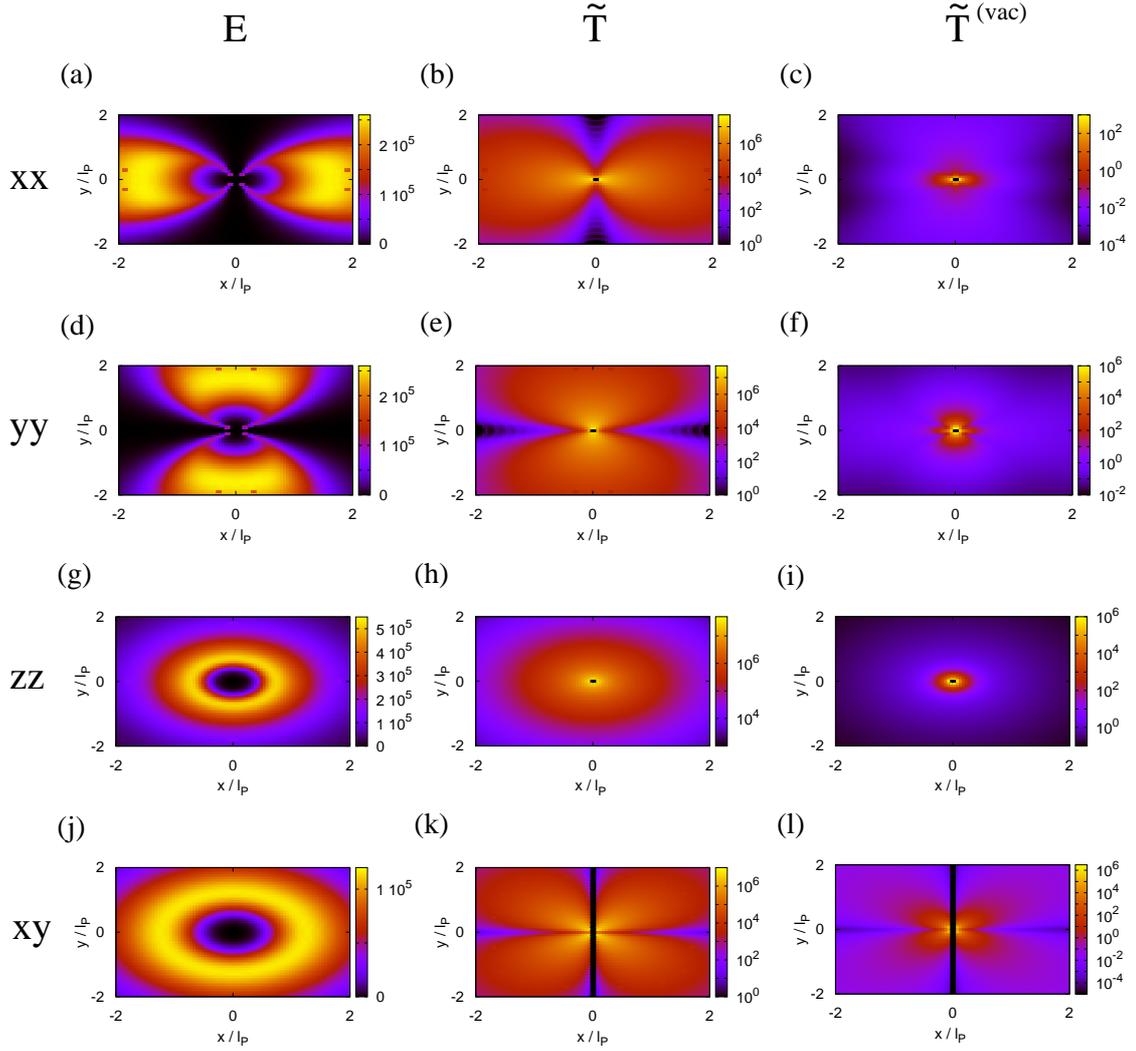, width = 0.9\textwidth}
  \caption{Enhancement factor $E$ (first column), the energy transfer function close to graphene $\tilde{T}$ (second column) and in free space $\tilde{T}^{(\rm vac)}$ (third column) in the  x-y plane for $\omega = 0.7\,{\rm eV}$ and (a)-(c) $\mathbf{e}_{\rm A} = \mathbf{e}_{\rm D} = \mathbf{e}_x$, (d)-(f)  $\mathbf{e}_{\rm A} = \mathbf{e}_{\rm D} = \mathbf{e}_y$, (g)-(i)  $\mathbf{e}_{\rm A} = \mathbf{e}_{\rm D} = \mathbf{e}_z$, and (j) -(l) $\mathbf{e}_{\rm A} = \mathbf{e}_x$ and $\mathbf{e}_{\rm D} = \mathbf{e}_y$. The x and y coordinates are normalized to the propagation length of the graphene plasmon $l_{\rm P} = 486\,{\rm nm}$. $z = 10\,{\rm nm}$ and $E_F = 1\,{\rm eV}$. \label{Fig:xyPlots}}
\end{figure}

In Fig.~\ref{Fig:xyPlots}(a),(d),(g),(j) we show plots of $E$ where the acceptor is placed at 
$\mathbf{r}_A = (0\,{\rm nm},0\,{\rm nm},10\,{\rm nm})$ and the x-y position of the donor position $\mathbf{r}_D = (x,y,10\,{\rm nm})$   is varied. Once more it becomes obvious that the largest enhancement factors can be obtained for distances 
around the propagation length of the graphene plasmons $l_{\rm P}$. In addition, we have plotted the energy
transfer function above graphene $\tilde{T}$ and in free space $\tilde{T}^{(\rm vac)}$.
Note that the enhancement factor $E$ is invariant under
rotation around the $z$ axis for $\mathbf{e}_{\rm A} = \mathbf{e}_x$ and $\mathbf{e}_{\rm D} = \mathbf{e}_y$
although $\tilde{T}$ and $\tilde{T}^{(\rm vac)}$ are not. 

\begin{figure}
  \epsfig{file = 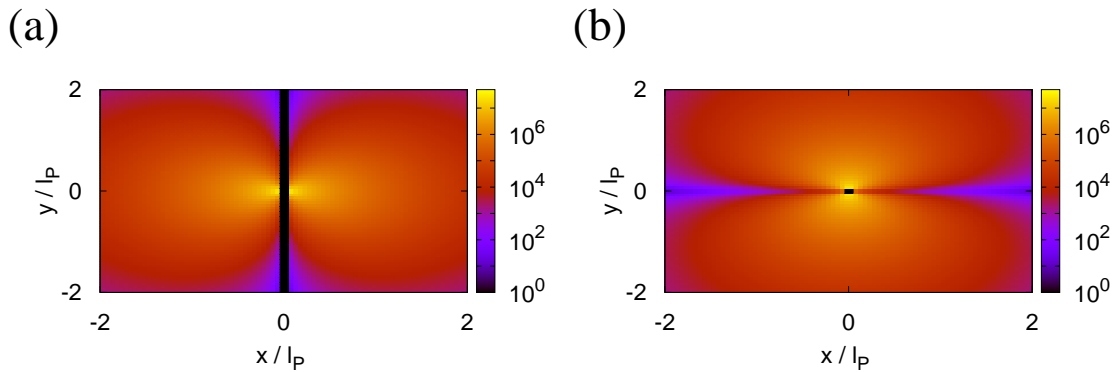, width = 0.9\textwidth}
  \caption{Energy transfer function $\tilde{T}$ for (a)  $\mathbf{e}_{\rm A} = \mathbf{e}_x$ and $\mathbf{e}_{\rm D} = \mathbf{e}_z$  and (b) $\mathbf{e}_{\rm A} = \mathbf{e}_y$ and $\mathbf{e}_{\rm D} = \mathbf{e}_z$ in x-y plane. The x and y axis are normalized to propagation length $l_{\rm P}$, $z = 10\,{\rm nm}$, $E_F = 1\,{\rm eV}$, and $\omega = 0.7\,{\rm eV}$. \label{Fig:crosscoupling}}
\end{figure}

It is interesting to note that due to the presence of the graphene sheet there is also FRET between the donor and
acceptor for ($\mathbf{e}_{\rm A} = \mathbf{e}_x$, $\mathbf{e}_{\rm D} = \mathbf{e}_z$) and  ($\mathbf{e}_{\rm A} = \mathbf{e}_y$,$\mathbf{e}_{\rm D} = \mathbf{e}_z$). The free space FRET vanishes in this case so that $E$ is not a well defined quantity for such orientations of dipole moments. But the energy transfer function $\tilde{T}(\omega)$ shown in Fig.~\ref{Fig:crosscoupling} indicate that
 the PE of FRET is in the same order of magnitude as for the cases considered in Fig.~\ref{Fig:xyPlots}.

\begin{figure}
  \epsfig{file = 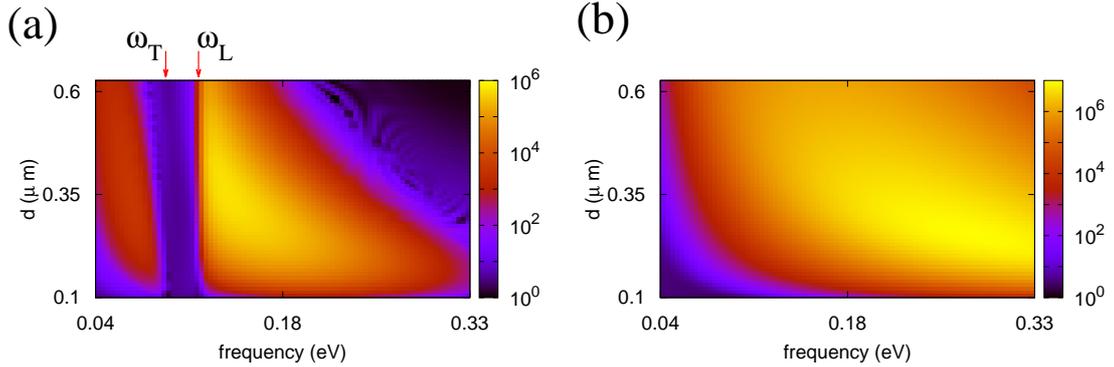, width = 0.9\textwidth}
  \caption{Enhancment factor $E$ (log scale) for a donor-acceptor pair above (a) graphene on a SiC substrate and (b) suspended 
           graphene as a function of 
           the donor-acceptor distance $d$ and the frequency; $z = 10\,{\rm nm}$ and $E_F = 0.5\,{\rm eV}$. The arrows mark
           the transversal and longitudinal phonon frequencies $\omega_{\rm T}$ and $\omega_{\rm L}$ of SiC.\label{Fig5}} 
\end{figure}

Finally, we briefly discuss the impact of the substrate. In Fig.~\ref{Fig5}(a) we show a plot of the enhancement factor $E$
of FRET above a graphene sheet deposited on SiC as a function of frequency and distance. We choose $z = 10\,{\rm nm}$ and
$E_F = 0.5\,{\rm eV}$. For comparison we also plot the enhancement factor for suspended graphene in Fig.~\ref{Fig5}(b)
using the same parameters (for similar plots using $E_F = 1\,{\rm eV}$ see Supporting Information). 
In contrast to the results found in Fig.~\ref{Fig5}(b) we can see 
a gap of negligible small FRET in the {\itshape reststrahlen} region of 
SiC (i.e.\ frequencies between the transversal and longitudinal phonon frequencies $\omega_{\rm T}$ and $\omega_{\rm L}$)~\cite{KochEtAl2012,MessinaEtAl2013}. In addition, due to the interaction of the graphene plasmon with the 
surface-phonon polaritons in SiC the PE is redshifted and concentrated at frequencies slighty larger 
than $\omega_{\rm L} = 0.12\,{\rm eV}$. It can also be seen that the enhancement factor on graphene/SiC is 
slightly smaller and the effect is more narrowband compared to the case of suspended graphene. 
Hence, by choosing an active or inactive substrate the 
observed FRET enhancement can be further controlled.

%
%

\section{Conclusion}

In summary we have shown that in close vicinity to a sheet of graphene FRET can be enhanced by
six orders of magnitude. The maximum enhancement is obtained for donor-acceptor distances on the
order of the propagation length of the plasmons in graphene. Further, we have shown that the enhancement
effect is broadband and depends not only on the relative position of the donor and acceptor but
also on the orientation of their dipole moments. In particular, the presence of graphene allows
for coupling of perpendicularly oriented dipole moments which do not couple in free space. The
role of the substrate is briefly discussed using SiC. Hence, the observed plasmonic enhancement
can be tuned by doping or gating of graphene and by choosing an appropriate substrate.


\end{document}